\documentclass[conference]{IEEEtran}
\IEEEoverridecommandlockouts

\usepackage[T1]{fontenc}
\usepackage{cite,amsmath,amssymb,amsthm,bm}
\usepackage{graphicx,booktabs,array,microtype}
\usepackage[hidelinks,hypertexnames=false]{hyperref}
\usepackage{algorithm,algpseudocode}
\makeatletter
\newcommand\fs@betterruled{%
  \def\@fs@cfont{\bfseries}\let\@fs@capt\floatc@ruled
  \def\@fs@pre{\vspace*{0.07in}\hrule height.8pt depth0pt \kern2pt}%
  \def\@fs@post{\kern2pt\hrule\relax}%
  \def\@fs@mid{\kern2pt\hrule\kern2pt}%
  \let\@fs@iftopcapt\iftrue}
\floatstyle{betterruled}
\restylefloat{algorithm}
\makeatother
\usepackage{flushend}
\def\BibTeX{{\rm B\kern-.05em{\sc i\kern-.025em b}\kern-.08em
    T\kern-.1667em\lower.7ex\hbox{E}\kern-.125emX}}

\newtheorem{proposition}{Proposition}
\newcommand{\E}{\mathbb E}
\newcommand{\Nch}{N_{\mathrm{ch}}}
\newcommand{\method}{SC-DJSCC}
\newcommand{\norm}[1]{\lVert #1\rVert}
\newcommand{\MCError}{0.0036}
\newcommand{\EntropyError}{$1.8\times10^{-15}$}

\begin{document}

\title{Synergy-Aware Symbol Allocation for Distributed Semantic Communication
\thanks{This work was supported in part by the National Natural Science Foundation of China under Grants 62595731 and 62293484, and the Program of Jiangsu Province under Grant NTACT-2024-Z-001.}}
\author{\IEEEauthorblockN{Fan Gao, Youzheng Wang, Zhijin Qin, and Xiaoming Tao}
\IEEEauthorblockA{Department of Electronic Engineering, BNRist, Tsinghua University, Beijing, China\\
gaof23@mails.tsinghua.edu.cn, \{yzhwang, qinzhijin, taoxm\}@tsinghua.edu.cn}}

\maketitle

\begin{abstract}
Distributed semantic communication enables spatially separated terminals to support collaborative inference over wireless links for 6G edge intelligence. In multimodal sensing and cooperative perception, a single terminal captures only a partial view, and full task semantics emerge only when complementary observations from multiple terminals are fused at the receiver. However, under tight uplink budgets, existing task-oriented joint source--channel coding (JSCC) and distributed information bottleneck (IB) methods evaluate rate--distortion trade-offs per terminal, tending to undervalue cross-terminal synergistic features and prune them prematurely. To address this issue, we propose a synergy-credited distributed JSCC (\method{}) framework that introduces a differentiable, post-equalization matrix R\'enyi synergy credit as a surrogate to reward complementary task evidence surviving the wireless channel and backpropagates its gradients to local symbol gates. At deployment, the credit branch is removed and binary symbol masks are frozen, allowing each terminal to encode independently without inter-terminal signaling or extra inference overhead. Experiments show that \method{} recovers the analytical symbol-selection threshold on parity sources and reduces payload symbols by 28.4\% on AV-MNIST and 32.9\% on PAMAP2 over an uncredited gated baseline, while achieving a 6.2-point accuracy gain at $-6$~dB under 3GPP TDL-C fading with pilot-aided channel estimation for low-overhead wireless cooperative sensing.
\end{abstract}
\begin{IEEEkeywords}
Semantic communication, distributed inference, joint source--channel coding, information bottleneck, symbol selection.
\end{IEEEkeywords}

\section{Introduction}
Distributed semantic communication supports collaborative inference across spatially separated terminals over wireless links~\cite{shao2023multidevice,gunduz2023beyond}. In multimodal sensing and cooperative perception, each terminal observes a partial view or modality and encodes it independently without inter-terminal coordination, while a fusion receiver combines the corrupted messages to reconstruct task semantics. Under strict uplink budgets, task-oriented joint source--channel coding (JSCC) must heavily compress local representations into a small number of transmitted channel symbols at each encoder~\cite{bourtsoulatze2019deepjscc,shao2022taskoriented}.

Conventional rate-constrained compression, however, tends to undervalue cross-terminal complementary features. By partial information decomposition (PID), multi-terminal observations contain redundant, unique, and \emph{synergistic information}, where synergistic semantics are uninformative at any single terminal alone and emerge only after fusion~\cite{williams2010pid,liang2023interactions}. If a binary label is the parity of two independent fair bits across two terminals, neither bit alone reduces label uncertainty, yet the pair determines the label. Because synergistic semantics require transmissions from multiple terminals simultaneously while yielding little standalone information at any single terminal, standard rate--distortion or information bottleneck (IB) objectives that evaluate rate--distortion trade-offs per terminal prune synergistic features prematurely as bandwidth tightens. Over wireless links, a complementary feature helps only if its partner transmissions from other terminals also survive fading and equalization.

Existing task-oriented JSCC and distributed IB formulations do not prevent this premature loss of synergistic semantics. Variable-length deep JSCC~\cite{bourtsoulatze2019deepjscc,shao2022taskoriented} and distributed IB methods~\cite{tishby2000ib,alemi2017vib,aguerri2021distributed} compress each terminal by penalizing its local rate term. Even when augmented with selective retransmission~\cite{shao2023multidevice}, subset-sufficiency constraints~\cite{zhou2025tadib}, or inter-terminal orthogonality regularizers~\cite{wang2026toib}, these methods still evaluate local representations marginally without rewarding complementary evidence upon fusion. Conversely, information-theoretic models of multimodal interactions~\cite{westphal2026gib,gao2026multimodal} operate on noiseless latent spaces rather than coupling post-equalization synergy returns to physical symbol gating.

\begin{figure*}[!t]
\centering\includegraphics[width=\textwidth]{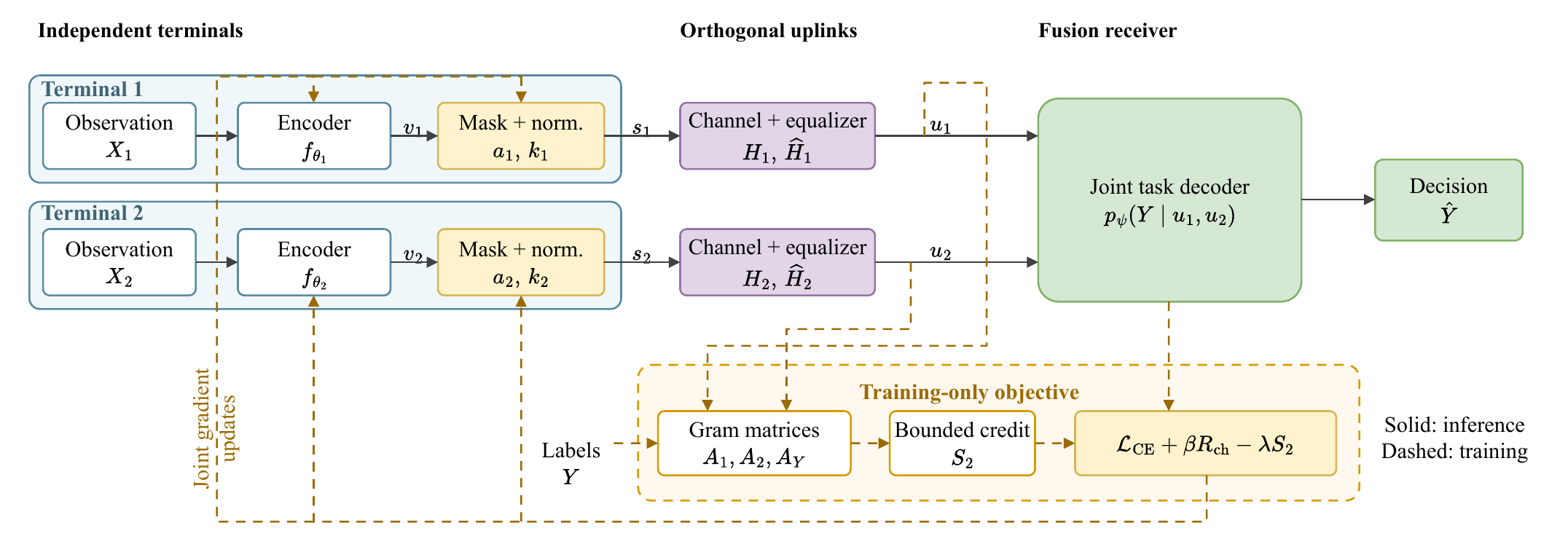}
\caption{SC-DJSCC with independent deployment terminals and a joint training objective. Solid arrows carry the inference messages. Dashed paths form Gram matrices from equalized messages and labels, compute the bounded credit, and update encoders, masks, and the fusion decoder during training only.}
\label{fig:system}
\end{figure*}

To address this issue, we propose a synergy-credited distributed JSCC (\method{}) framework that introduces a post-equalization synergy credit as a surrogate during training with zero inference overhead at deployment:
\begin{itemize}
\item \textbf{Synergy-Credited Distributed JSCC Framework:} We propose \method{}, coupling local symbol gates with a training-only, post-equalization synergy credit and removing the credit branch at deployment for zero inference or signaling overhead. Closed-form thresholds prove it corrects the $1/K$ pruning bias against $K$-terminal complementary features according to link reliability.
\item \textbf{Post-Equalization Matrix R\'enyi Synergy Surrogate:} We formulate a bounded order-two matrix R\'enyi synergy credit as a surrogate on equalized messages and backpropagate its gradients to local gates, rewarding super-additive joint task evidence in $O(B^2)$ Frobenius-norm operations.
\item \textbf{Experimental Validation:} Controlled parity tests show that \method{} tracks the synergy fraction and recovers the analytical selection threshold, while evaluations on AV-MNIST and PAMAP2 demonstrate substantial symbol savings and enhanced robustness under 3GPP TDL-C fading.
\end{itemize}

\section{System Model}
\label{sec:system}
\subsection{Transmission and Receiver Model}
We consider a two-terminal distributed semantic communication system supporting a shared task $Y$ from observations $X_1,X_2$ over orthogonal links (Fig.~\ref{fig:system}). Following analog deep JSCC~\cite{bourtsoulatze2019deepjscc}, encoder $m$ maps $X_m$ directly to $L_m$ candidate complex baseband symbols $v_m=f_{\theta_m}(X_m)\in\mathbb C^{L_m}$, with real and imaginary parts forming the in-phase and quadrature components, and applies a binary selection mask $a_m\in\{0,1\}^{L_m}$. With $k_m=\norm{a_m}_0$ active channel uses, the transmitted signal is
\begin{equation}
 s_m=\sqrt{k_mP_m}\frac{a_m\odot v_m}{\norm{a_m\odot v_m}_2},\qquad
 \Nch=k_1+k_2 .\label{eq:tx}
\end{equation}
If the masked vector is zero, $s_m=0$; otherwise its energy is $k_mP_m$. Only active entries are transmitted, so a closed gate saves a complex channel use rather than sending zero. The bandwidth resource is $\Nch$, and the channel-bandwidth ratio (CBR) is $\Nch/D_{\rm src}$, where $D_{\rm src}$ counts the preprocessed real source scalars.

The channel and receiver are
\begin{equation}
 r_m=H_ms_m+n_m,\quad u_m=\mathcal E(r_m,\widehat H_m),\quad
 p_\psi(Y\mid u_1,u_2),\label{eq:channel}
\end{equation}
where $n_m\sim\mathcal{CN}(0,N_0I)$ and $\mathcal E$ is a fixed equalizer using receiver channel-state information (CSI). AWGN uses $H_m=I$, whereas OFDM over multipath fading has diagonal $H_m$ after cyclic-prefix removal and FFT. With per-subcarrier LMMSE equalization,
\begin{equation}
 u_{m,k}=\frac{\widehat H_{m,k}^{*}r_{m,k}}{|\widehat H_{m,k}|^2+N_0/P_m}.
 \label{eq:equalizer}
\end{equation}
Pilot and cyclic-prefix overheads are identical across all compared schemes, so $\Nch$ counts payload symbols only. Training uses paired observations and labels centrally; at deployment, terminal $m$ uses only $X_m$, masks are conveyed once per operating point, and the receiver restores transmitted entries to their agreed positions before fusion.

\subsection{Complementary-Coordinate Selection}

We next analyze how joint-only task evidence alters the per-symbol selection threshold over noisy links. Let each retained coordinate cost one channel use. For a shared fair bit, either terminal can deliver one bit of task relevance. For parity, $Y=X_1\oplus X_2$ with independent fair inputs, both coordinates must be delivered. Consider the coordinate-selection objective
\begin{equation}
 \mathcal J=-I(Y;U_1,U_2)+\beta(k_1+k_2)-\lambda S,\label{eq:coordinate}
\end{equation}
where $\beta,\lambda\geq0$, and $S$ is zero for inactive or singly informative choices and equals $\kappa\in[0,1]$ for the retained complementary pair.

\begin{proposition}[Complementary-coordinate selection]
\label{prop:threshold}
Let $X_1,X_2$ pass through independent binary symmetric channels with crossover probabilities $e_1,e_2\leq1/2$. The received pair supplies
\begin{equation}
 J(e_1,e_2)=1-H_{\rm b}(e_1+e_2-2e_1e_2)\label{eq:parity}
\end{equation}
bits about $Y$, where $H_{\rm b}(\cdot)$ is the binary entropy function, while either received input alone supplies zero. Among the four coordinate masks, retaining both is strictly preferred in~\eqref{eq:coordinate} when
\begin{equation}
 \beta<\frac{J(e_1,e_2)+\lambda \kappa}{2}.\label{eq:threshold}
\end{equation}
\end{proposition}
\begin{proof}
The parity error occurs exactly when one channel flips and the other does not, giving crossover $e_1+e_2-2e_1e_2$. Symmetry makes the received parity a sufficient statistic for $Y$ and makes each single input independent of $Y$. The empty and single-coordinate choices have objectives $0,\beta,\beta$; the pair has $-J+2\beta-\lambda \kappa$. Comparing these four values proves the result.
\end{proof}

\begin{figure}[!t]
\centering\includegraphics[width=\columnwidth]{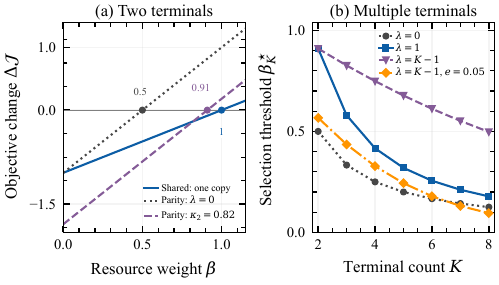}
\caption{Coordinate-model analysis. (a) Noiseless two-terminal objective changes with calibrated credit $\kappa_2=0.82$ and $\lambda=1$. (b) $K$-terminal thresholds with calibration-matched population Gaussian kernels, $c_m\approx0.258$ and $\epsilon\to0$; the noisy curve uses $e_m=0.05$, the others $e_m=0$.}
\label{fig:selection}
\end{figure}

Without credit, noiseless parity is retained only for $\beta<1/2$, whereas a shared bit is retained for $\beta<1$ since parity requires active symbols from both terminals. Fig.~\ref{fig:selection}(a) uses the calibrated credit $\kappa_2=0.82$, which raises the pair's threshold to $0.91$ at $\lambda=1$ while accounting for both symbols. Channel corruption lowers $J$ in~\eqref{eq:parity} and the threshold with it, so link reliability enters symbol selection directly.

\paragraph{Extension to multiple terminals}
For $K$ independent fair inputs with $Y=\bigoplus_{m=1}^K X_m$, every proper subset of received inputs is independent of $Y$. Under independent binary symmetric channels, the full set supplies
\begin{equation}
 J_K=1-H_{\rm b}(q_K),\qquad
 q_K=\frac{1-\prod_{m=1}^K(1-2e_m)}{2}.
 \label{eq:parityK}
\end{equation}
With zero credit for proper subsets and full-set credit $\kappa_K\in[0,1]$, comparing all $2^K$ masks shows that the full set is strictly preferred when $\beta<\beta_K^\star\triangleq(J_K+\lambda \kappa_K)/K$. Without credit, the threshold decays as $1/K$, so cost-aware selection increasingly drops higher-order complementary evidence as more terminals participate; $\kappa_K$ is the bounded group credit defined in Section~\ref{sec:method}.

For a population binary-kernel reference, let $c_m$ be the Gaussian-kernel similarity between the two received bit values and $\mu_K=1-2q_K$. The order-two credit is
\begin{equation}
 \kappa_K=\frac{\log_2(1+\mu_K^2\prod_m \xi_m)}{1+\epsilon},\qquad
 \xi_m=\frac{1-c_m^2}{1+c_m^2}.
 \label{eq:creditKreference}
\end{equation}
With $C_m=(1+c_m^2)/2$ and $D_m=(1-c_m^2)/2$, the joint message--label collision trace is $(\prod_m C_m+\mu_K^2\prod_m D_m)/2$; each single message is label-independent, giving~\eqref{eq:creditKreference}. Fig.~\ref{fig:selection}(b) fits $c_m\approx0.258$ to $\kappa_2=0.82$ at $K=2$ ($\epsilon\to0$). Fixed $\lambda=1$ adds $\kappa_K/K$ to the threshold, whereas $\lambda=K-1$ offsets the $1/K$ decay and recovers unit threshold as $e_m,c_m,\epsilon\to0$.

\section{Synergy-Credited Distributed JSCC}
\label{sec:method}
\subsection{Differentiable Symbol Selection}
To select active channel symbols per terminal end-to-end, we learn sample-independent mask logits $\eta_{mk}$ via the hard-concrete relaxation~\cite{louizos2018l0}. With $\nu\sim\mathcal U(0,1)$,
\begin{align}
 t_{mk}&=\operatorname{sigmoid}\!\left(\frac{\log \nu-\log(1-\nu)+\eta_{mk}}{\tau_g}\right),\\
 a_{mk}&=\operatorname{clip}_{[0,1]}\big(\gamma+(\zeta-\gamma)t_{mk}\big),
\end{align}
where $\gamma<0<1<\zeta$. The expected nonzero count is
\begin{equation}
 R_{\rm ch}=\sum_{m,k}\pi_{mk},\quad
 \pi_{mk}=\operatorname{sigmoid}\!\left(\eta_{mk}-\tau_g\log\frac{-\gamma}{\zeta}\right).
 \label{eq:rate}
\end{equation}
During relaxed training, the power normalization uses the differentiable effective count $\sum_k a_{mk}$ in place of $k_m$. Deployment thresholds the deterministic stretched gates at $1/2$ and reapplies~\eqref{eq:tx} with the resulting integer count. Each rate weight defines a separately trained operating point.

\subsection{Post-Channel Interaction Credit}
To reward complementary evidence surviving the channel, we evaluate a bounded credit on the equalized messages. For a batch of $B$ paired examples, we realify each equalized complex message and build a Gaussian Gram matrix with bandwidth $\omega_m$:
\begin{equation}
 [G_m]_{ij}=\exp\!\left(-\frac{\norm{u_{m,i}-u_{m,j}}_2^2}{2\omega_m^2}\right),
 \quad A_m=G_m/\operatorname{tr}G_m .
\end{equation}
For labels, $[G_Y]_{ij}=\mathbf1\{y_i=y_j\}$ and $A_Y=G_Y/B$. All matrices are positive semidefinite with unit trace. Joint relations use the normalized Hadamard product
\begin{equation}
 A\star C=\frac{A\odot C}{\operatorname{tr}(A\odot C)} .\label{eq:joint}
\end{equation}
The matrix-based R\'enyi entropy is~\cite{sanchezgiraldo2015entropy,yu2020multivariate}
\begin{align}
 H_\alpha(A)&=\frac{\log_2\operatorname{tr}(A^\alpha)}{1-\alpha},\\
 I_\alpha(A;C)&=H_\alpha(A)+H_\alpha(C)-H_\alpha(A\star C).
\end{align}
Define $A_{12}=A_1\star A_2$ and the task-conditioned interaction excess
\begin{equation}
 \Delta_\alpha=I_\alpha(A_{12};A_Y)-I_\alpha(A_1;A_Y)-I_\alpha(A_2;A_Y).
 \label{eq:delta}
\end{equation}
With $\epsilon>0$, the bounded credit is
\begin{equation}
 S_\alpha=\min\!\left\{1,\frac{[\Delta_\alpha]_+}{H_\alpha(A_Y)+\epsilon}\right\},
 \qquad [x]_+=\max(0,x).\label{eq:credit}
\end{equation}
A single-class batch sets $S_\alpha=0$. The credit is zero for nonpositive excess, bounded in $[0,1]$, and differentiable almost everywhere.

For $K$ terminals, set $A_{[K]}=A_1\star\cdots\star A_K$ and use
\begin{equation}
 \Delta_{\alpha,K}=I_\alpha(A_{[K]};A_Y)-\sum_{m=1}^K I_\alpha(A_m;A_Y).
 \label{eq:deltaK}
\end{equation}
Applying~\eqref{eq:credit} preserves the same task normalization and bound for the full-group credit.

In bivariate PID, the Shannon counterpart of~\eqref{eq:delta} is interaction information, equal to synergy minus redundancy~\cite{williams2010pid}. Thus, $S_\alpha$ does not isolate pure PID synergy when strong redundancy drives $\Delta_\alpha\le0$; however, redundancy-dominated features already receive sufficient single-terminal task gradients in~\eqref{eq:objective} and need no rate compensation, so $[\Delta_\alpha]_+$ selectively targets the synergy-dominated regime where $\Delta_\alpha>0$ and per-symbol gating prunes complementary coordinates. Although kernel estimators are sensitive to feature scaling, bandwidth mismatch, and small batches, power normalization~\eqref{eq:tx}, detached batch-median bandwidths, and moderate batches keep $A_m$ well-conditioned and prevent scale manipulation.

\subsection{Training and Implementation}
To avoid cubic eigendecomposition during training, we evaluate the credit at order $\alpha=2$ and optimize it jointly with the task and rate terms. For a real symmetric trace-one matrix,
\begin{equation}
 H_2(A)=-\log_2\operatorname{tr}(A^2)=-\log_2\norm{A}_F^2.\label{eq:fast}
\end{equation}
Evaluating the entropy terms in~\eqref{eq:delta} then reduces to entrywise products, sums, and logarithms, costing $O(B^2)$ time and memory after Gram construction. Kernel construction costs $O(B^2d_u)$ for received feature width $d_u$, whereas an eigendecomposition would cost $O(B^3)$.

The training objective is
\begin{equation}
 \mathcal L=\E[-\log p_\psi(Y\mid u_1,u_2)]+\beta R_{\rm ch}-\lambda S_2.
 \label{eq:objective}
\end{equation}
For a gate logit, the three objective terms give
\begin{align}
 \frac{\partial\mathcal L}{\partial\eta_{mk}}
 &=\frac{\partial\mathcal L_{\rm task}}{\partial\eta_{mk}}
   +\beta\pi_{mk}(1-\pi_{mk})
   -\lambda\frac{\partial S_2}{\partial\eta_{mk}}.
 \label{eq:gategradient}
\end{align}
Task and credit gradients propagate through the channel, power normalization, and encoder, retaining symbols whose direct or complementary return outweighs their rate cost.

Algorithm~\ref{alg:train} uses three training phases: task warmup, increasing $\beta$ with $\lambda=0$, and full-objective optimization. Deployment freezes binary masks and removes the credit branch and labels, so inference cost equals that of the matched backbone at the same symbol count.

\begin{algorithm}[t]
\caption{Joint training, independent deployment}\label{alg:train}
\begin{algorithmic}[1]
\For{each paired training batch}
\State Encode locally, sample gates, normalize active-symbol power.
\State Simulate links and equalize received messages.
\State Form $A_1,A_2,A_Y$ and evaluate $S_2$ by~\eqref{eq:fast}.
\State Update encoders, masks, and decoder with~\eqref{eq:objective}.
\EndFor
\State Freeze binary masks; record $\Nch$ and configure receiver positions.
\State Deploy local encoders and fusion decoder; remove the credit branch.
\end{algorithmic}
\end{algorithm}

\begin{figure}[!t]
\centering\includegraphics[width=\columnwidth]{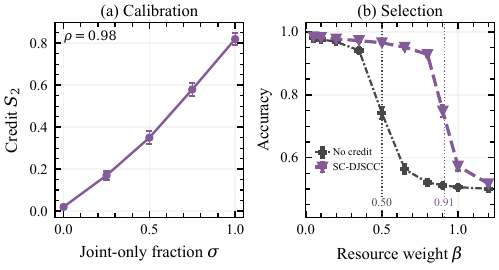}
\caption{Controlled mechanism verification over three random seeds. (a) Credit versus known joint-only fraction (Spearman $\rho$ over 15 trials). (b) Pure-parity accuracy versus resource weight; vertical lines mark analytical predictions $0.50$ and $0.91$, and error bars denote standard deviations.}
\label{fig:controlled}
\end{figure}

\begin{figure*}[!t]
\centering\includegraphics[width=\textwidth]{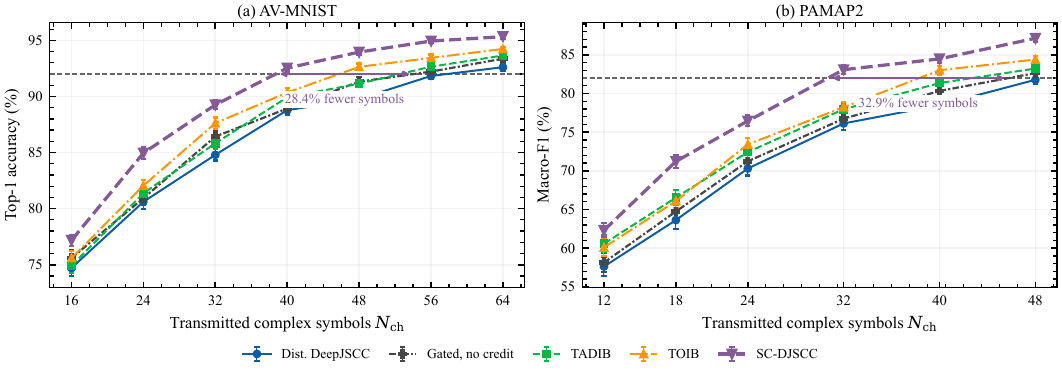}
\caption{AWGN task--symbol curves at 0 dB (mean and standard deviation over three random seeds). Horizontal lines indicate 92\% AV-MNIST accuracy and 82\% PAMAP2 macro-F1; arrows indicate the payload-symbol savings of SC-DJSCC over the gated control at these targets.}
\label{fig:pareto}
\end{figure*}

\begin{figure}[!t]
\centering\includegraphics[width=\columnwidth]{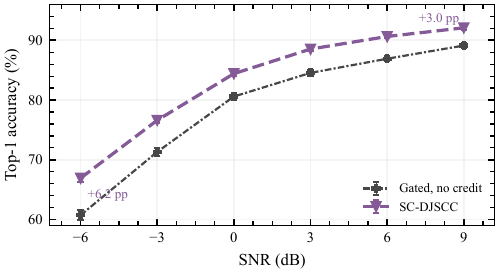}
\caption{AV-MNIST under TDL-C fading with comb-pilot LMMSE channel estimation at 40 payload symbols without retraining. The gain over the gated control rises from 3.0 points at 9 dB to 6.2 points at $-6$ dB.}
\label{fig:robust}
\end{figure}

\section{Experimental Results}
\label{sec:results}
\subsection{Experimental Setup}
We evaluate \method{} on two distributed multimodal benchmarks over AWGN and multipath fading channels. AV-MNIST~\cite{liang2021multibench} places image and audio observations at separate terminals and reports ten-class top-1 accuracy, using a LeNet-style encoder and complex baseband projection per modality. PAMAP2~\cite{reiss2012pamap2} places wrist IMUs at one terminal and chest--ankle IMUs at the other, using one-dimensional CNN encoders and macro-F1 over 12 activities; subjects 1--6/7/8--9 form train/validation/test splits before 2.56-s windows at 100 Hz with 50\% overlap, giving $D_{\rm src}=13328$ and $6912$. All schemes share the two-layer fusion decoder, normalization, and paired channel realizations, and are trained per budget over three seeds:
\begin{itemize}
\item \textit{Fixed-budget DeepJSCC}~\cite{bourtsoulatze2019deepjscc}: uses a fixed equal split $k_1=k_2=\Nch/2$ without masks, trained with task loss only.
\item \textit{Gated control with $\lambda=0$}: uses the same hard-concrete symbol masks and rate penalty $\beta R_{\rm ch}$, without interaction credit.
\item \textit{TADIB}~\cite{zhou2025tadib}: task-aware distributed IB, which compresses per-terminal rates while preserving modality-subset relevance.
\item \textit{TOIB}~\cite{wang2026toib}: task-oriented orthogonalized IB, which combines distributed IB compression with an inter-terminal orthogonality penalty to reduce redundancy.
\end{itemize}

The default configuration is $L_m=128$, $B=128$, $(\gamma,\zeta,\tau_g)=(-0.1,1.1,2/3)$, $\alpha=2$, and $\lambda=1$. Adam starts at $10^{-3}$ with cosine decay. Training SNR is sampled over $[-6,12]$ dB, and main task--symbol comparisons use AWGN at 0 dB. Multipath evaluations reuse the trained masks and weights without retraining over 3GPP TDL-C fading~\cite{3gpp38901} with a 300-ns delay spread using OFDM with 128 subcarriers, 30-kHz spacing, a 16-sample cyclic prefix, and comb-pilot LMMSE channel estimation.

\subsection{Credit Calibration and Threshold Verification}
We first verify that the interaction surrogate tracks known synergistic structure and reproduces the analytical selection threshold. Over 200,000 binary symmetric channel realizations per operating point across five seeds with $e_1\in\{0,0.1\}$ and $e_2\in[0,1/2]$, empirical mutual information matches~\eqref{eq:parity} within \MCError{} bits, and all 3,000 noiseless coordinate-mask decisions agree with Proposition~\ref{prop:threshold} away from ties. For Gaussian Gram matrices with $B\in\{16,32,64,128\}$, the Frobenius identity~\eqref{eq:fast} matches eigendecomposition within \EntropyError{}.

On a 32-coordinate source mixing shared and parity components in a known joint-only fraction $\sigma$, Fig.~\ref{fig:controlled}(a) shows that $S_2$ rises monotonically from $0.02$ at $\sigma=0$ to $0.82$ at $\sigma=1$ with Spearman $\rho=0.98$ over 15 trials. At $\sigma=0$, marginal alignments $I_2(A_m;A_Y)$ account for $I_2(A_{12};A_Y)$, so $\Delta_2\le0$ and the rectifier in~\eqref{eq:credit} suppresses the credit; as $\sigma$ grows, $A_1\star A_2$ aligns with $A_Y$ only when both views are combined, generating super-additive excess $\Delta_2>0$ proportional to $\sigma$.

On pure parity with $\sigma=1$, raising $\beta$ in Fig.~\ref{fig:controlled}(b) causes a sharp drop to chance accuracy once the rate penalty exceeds the task return. Because closing either gate destroys the parity statistic, uncredited gates with $\lambda=0$ cannot resist the rate penalty $\beta\pi_{mk}(1-\pi_{mk})$ in~\eqref{eq:gategradient} beyond $\beta\approx0.50$. Activating the credit with $\lambda=1$ injects the restoring gradient $-\lambda\partial S_2/\partial\eta_{mk}$ while both gates are open, shifting the retention cliff to $\beta\approx0.90$ and matching $(1+\lambda\kappa_2)/2=0.91$.

\subsection{Task Quality and Payload Efficiency}
We next examine how gate-gradient compensation shapes task quality across payload budgets in Fig.~\ref{fig:pareto}. At large budgets where $\beta$ is small, even uncredited gates retain both standalone and complementary features. As the budget tightens, gates prune coordinates in ascending order of their standalone marginal gradient $\partial\mathcal L_{\rm task}/\partial\eta_{mk}$; because cross-modal complementary features require partner symbols from both links simultaneously, their per-symbol return is halved and the control prunes them first. TADIB and TOIB also degrade under tight budgets: TADIB enforces subset sufficiency, favoring self-contained single-terminal features, while TOIB penalizes inter-terminal similarity unconditionally without separating task-relevant synergy from task-irrelevant orthogonal variations. Conditioning the post-channel excess on $A_Y$ protects coordinates whose joint kernel structure separates classes, yielding peak gains of $3.6$ points at 40 AV-MNIST symbols and $6.3$ points at 32 PAMAP2 symbols.

For a target quality $a_0$, let $\Nch^{(b)}(a_0)$ denote the piecewise-linear interpolation of the mean curve of scheme $b$ at $a_0$, with payload saving
\begin{equation}
 G_{\rm ch}(a_0)=1-\frac{\Nch^{\rm method}(a_0)}{\Nch^{\rm control}(a_0)}.
 \label{eq:saving}
\end{equation}
At the target lines in Fig.~\ref{fig:pareto}, \method{} cuts required symbols from 54.1 to 38.7 on AV-MNIST and from 45.8 to 30.7 on PAMAP2 over the control, saving 28.4\% and 32.9\% while outperforming TADIB and TOIB. Among discrete budgets, \method{} hits both targets at 40 and 32 symbols versus 56 and 48 for the control, while fixed-budget DeepJSCC needs 64 on AV-MNIST and fails to reach the PAMAP2 target.

Mechanistically, these savings arise from rebalancing active symbols across terminals. Audio on AV-MNIST and the wrist IMU on PAMAP2 act as auxiliary modalities that provide disambiguating cues only when fused with the dominant modality. Under uncredited rate compression with $\lambda=0$, standalone gradients dominate~\eqref{eq:gategradient} and under-allocate symbols to the auxiliary terminal, leaving 26/14 and 9/23 splits at $\Nch=40$ and $32$. Because $S_2$ depends on $A_1\star A_2$, its gradient vanishes if either terminal drops partner coordinates, shifting the splits to 22/18 and 14/18 and raising the fusion gain over the best single-terminal score from 5.8 to 8.6 points on AV-MNIST and from 6.8 to 11.9 points on PAMAP2.

\subsection{Ablation Study and Channel Robustness}
Finally, we analyze the interplay between post-channel evaluation, gating, and channel impairments. In Table~\ref{tab:ablation}, computing the credit before the channel on noiseless encoder outputs $v_m$ drops accuracy from 92.5\% to 90.6\%, as the gates retain source-complementary coordinates even when link noise degrades their post-equalization return $J(e_1,e_2)$ in~\eqref{eq:threshold}. Fixing uniform masks with $k_1=k_2=20$ under post-channel credit drops accuracy to 90.1\%, since the credit regularizes encoder weights $\theta_m$ without reallocating channel uses via~\eqref{eq:gategradient}.

\begin{table}[!t]
\caption{AV-MNIST ablation at 0 dB and 40 complex symbols. Mean$\pm$standard deviation over three random seeds.}
\label{tab:ablation}\centering
\begin{tabular}{@{}lc@{}}\toprule
Training configuration & Accuracy (\%)\\\midrule
No interaction credit & $88.9\pm0.4$ \\
Credit before channel & $90.6\pm0.4$ \\
Credit with fixed symbol masks & $90.1\pm0.5$ \\
SC-DJSCC & $\mathbf{92.5\pm0.3}$ \\
\bottomrule\end{tabular}
\end{table}

When tested zero-shot over 3GPP TDL-C multipath fading with comb-pilot LMMSE channel estimation in Fig.~\ref{fig:robust}, the gain of \method{} over the control more than doubles from 3.0 points at 9 dB to 6.2 points at $-6$ dB. Under frequency-selective fading and CSI errors, LMMSE equalization~\eqref{eq:equalizer} leaves structured residual noise that perturbs pairwise distances in $A_m$. Because maximizing $S_2$ during variable-SNR training requires sample pairs to preserve within-class kernel similarity simultaneously in $A_1$ and $A_2$, backpropagating $\partial S_2/\partial u_m$ drives encoders to pack complementary cues into larger-margin, fade-resilient subspace directions rather than fragile fine-grained coordinates.

Sensitivity sweeps peak at $\lambda=1$ and $\alpha=2$: smaller $\lambda$ under-compensates the two-symbol cost, while larger $\lambda$ over-weights interaction relative to marginal sufficiency. Discarding the credit branch after training leaves deployment identical to the gated backbone at 0.42 M parameters, 12.8 MMACs, and 0.41-ms batch-one latency on an RTX 4090.

\section{Conclusion}
We proposed \method{} to prevent per-terminal rate--distortion compression from prematurely pruning cross-terminal synergistic features in distributed semantic communication. Coupling local symbol gates with a training-only, post-equalization matrix R\'enyi synergy credit as a surrogate restores the retention threshold of complementary coordinates according to link reliability with zero inference or signaling overhead at deployment. Experiments showed that \method{} recovers the analytical threshold on parity sources and cuts payload symbols by 28.4\% on AV-MNIST and 32.9\% on PAMAP2, with gains reaching 6.2 points at $-6$~dB under TDL-C fading. While our model extends to $K>2$ terminals, real-world tests here cover $K=2$, leaving $K>2$ validation to future work.
\bibliographystyle{IEEEtran}
\bibliography{IEEEabrv,ref}
\end{document}